\documentclass[aps,prl,preprint,groupedaddress]{revtex4-1}

\usepackage{amsmath}
\usepackage{graphicx}
\usepackage{epstopdf}

\begin{document}


\title{Cavity mediated coherent coupling of magnetic moments}


\author{N. J. Lambert}
\affiliation{Microelectronics Group, Cavendish Laboratory, University of Cambridge, Cambridge, CB3 0HE, UK}

\author{J. A. Haigh}
\affiliation{Hitachi Cambridge Laboratory, Cavendish Laboratory, University of Cambridge, Cambridge, CB3 0HE, UK}

\author{S. Langenfeld}
\affiliation{Microelectronics Group, Cavendish Laboratory, University of Cambridge, Cambridge, CB3 0HE, UK}

\author{A. C. Doherty}
\affiliation{Centre for Engineered Quantum Systems, School of Physics, University of Sydney, Sydney, NSW 2006, Australia}

\author{A. J. Ferguson}
\email[]{ajf1006@cam.ac.uk}
\affiliation{Microelectronics Group, Cavendish Laboratory, University of Cambridge, Cambridge, CB3 0HE, UK}


\date{\today}

\begin{abstract}
We demonstrate the long range strong coupling of magnetostatic modes in spatially separated ferromagnets mediated by a microwave frequency cavity. Two spheres of yttrium iron garnet are embedded in the cavity and their magnetostatic modes probed using a dispersive measurement technique. We find they are strongly coupled to each other even when detuned from the cavity modes. We investigate the dependence of the magnet-magnet coupling on the cavity detuning $\Delta$, and find a $1/\Delta$ dependence also characteristic of cavity-coupled superconducting qubits. Dark states of the coupled magnetostatic modes of the system are observed, and ascribed to mismatches between the symmetries of the modes and the drive field.
\end{abstract}


\maketitle

There are several mechanisms by which spin angular momenta can couple, producing new total spin eigenstates. In most circumstances, the dominant coupling is either the dipole interaction, where the magnetic dipole moments interact through the electromagnetic field, or exchange coupling through a combination of the wavefunction symmetry and electrostatic interactions. The coupling of spins does not always occur directly; in many cases indirect mechanisms play an important role. The additional element in an exchange-type interaction can be a localised electron as in super-exchange, or an itinerant carrier as in the RKKY interaction\cite{Blundell2001}, which enables magnetic coupling between two ferromagnets through a thin electrically conducting paramagnetic spacer layer\cite{Parki1990,Bruno1991,Parkin1991}. The dipole interaction can also be involved in indirect coupling, such as in the J-coupling of nuclear spins mediated by a combination of the local dipole coupling between the nuclear spin and electrons and exchange coupling of the electrons on the separate nuclei\cite{Ramsey1952}.

An alternative indirect coupling between spin-like objects has been demonstrated in the context of cavity\cite{Haroche2006} and circuit\cite{Blais2004,Wallraff2004} QED; in this case the additional element is a low-loss resonator. It is not necessary for the psuedospins to be resonant with the cavity. Instead the spins and cavity can be significantly detuned from one another, in what is known as the dispersive regime. Here, an interaction between the pseudospins and the cavity via the local dipole coupling leads to an effective coupling between the individual pseudospins mediated by virtual photons\cite{Blais2004,Blais2007}. This approach can be used to couple together quantum systems such as qubits in a controlled way at distances far beyond that achieved by the qubit's dipole alone\cite{Majer2007,Sillanpaa2007,McKay2015}. The versatility of this approach is demonstrated by the variety of alternative systems which can be coupled to the resonator, including spin ensembles\cite{Schuster2010,Amsuss2011,Simmons2011,Putz2014}, double quantum dots\cite{Frey2012,Petersson2012} and hybrid systems\cite{Zhu2011,Marcos2010}.

Here we apply a dispersive technique to demonstrate long-range coupling of the macroscopic magnetic moments of two ferrimagnetic yttrium iron garnet (YIG) spheres mediated by an electromagnetic cavity. Due to its low damping YIG is an important material for microwave components such as tunable filters and couplers\cite{Helszajn1985}, as well as current research into spintronics\cite{Kajiwara2010,Qu2013,Nakayama2013}. Strong coupling between YIG magnetostatic modes and microwave cavities is readily attainable\cite{Zhang2014,Huebl2013,Goryachev2014a,Lambert2015}, and has been exploited in recent work\cite{Tabuchi2014} towards coherent control of single magnon states. In such systems changes in longitudinal magnetisation shift the cavity mode frequency\cite{Haigh2015}, yielding a dispersive measurement of ferromagnetic resonance (FMR).

In our experiments the two 1 mm diameter YIG spheres are placed within the dielectric of a co-axial transmission line cavity, made by breaking the inner of a 3.5 mm semirigid coaxial
cable in two places. They are positioned at the antinodes of the second harmonic resonance ($\omega_2/2\pi=7.15$ GHz, loss rate $\kappa/2\pi=45$ MHz) (Fig.~1a). A magnetic field is applied in the coaxial direction, consisting of a uniform field, $H_0$, and a differential field, $\pm\delta H/2$, local to each sphere. $H_0$ is sufficiently large to saturate the magnetisation of the YIG. In addition to the uniform magnetostatic mode of linewidth $\nu/2\pi\approx5$ MHz, a single domain ferromagnetic sphere hosts a spectrum of non-uniform magnetostatic modes which lie in a frequency band around the uniform mode\cite{Walker1957,Fletcher1959}. We focus our attention on the uniform mode, corresponding to the magnetisation precessing in phase throughout the sphere; this couples most strongly to the resonator field.

In order to characterise the strength of the resonant magnet-cavity coupling we measure the transmission of a -10 dBm probe tone of frequency $\omega_p/2\pi$ through the cavity. Initially we detune the modes of the two spheres from each other by $\approx450$ MHz ($\delta H \approx 16$ mT) and measure the cavity transmission as a function of $\omega_p$ and $H_0$ around its second harmonic (Fig.~1b). The FMR frequencies of the spectrum of magnetostatic modes are seen to come separately into resonance with the microwave cavity and the avoided crossings show that the uniform modes of the YIG spheres are strongly coupled to the cavity\cite{Zhang2014,Goryachev2014a,Tabuchi2014,Huebl2013}. For this cavity mode we determine a magnet-cavity coupling of $g_{\omega_2}/2 \pi\approx 150$ MHz for both spheres, and we find the coupling to higher order magnetostatic modes to be weaker\cite{Lambert2015}. Here, $\frac{g^2_{\omega_2}}{\kappa \nu}\approx 100$, confirming that the system is in the strong coupling regime. Similar behaviour (not shown) occurs around the fundamental mode of the cavity, $\omega_1/2 \pi=3.55$ GHz, with $g_{\omega_1}/2 \pi \approx 80$ MHz.

\begin{figure}
\begin{center}
\includegraphics{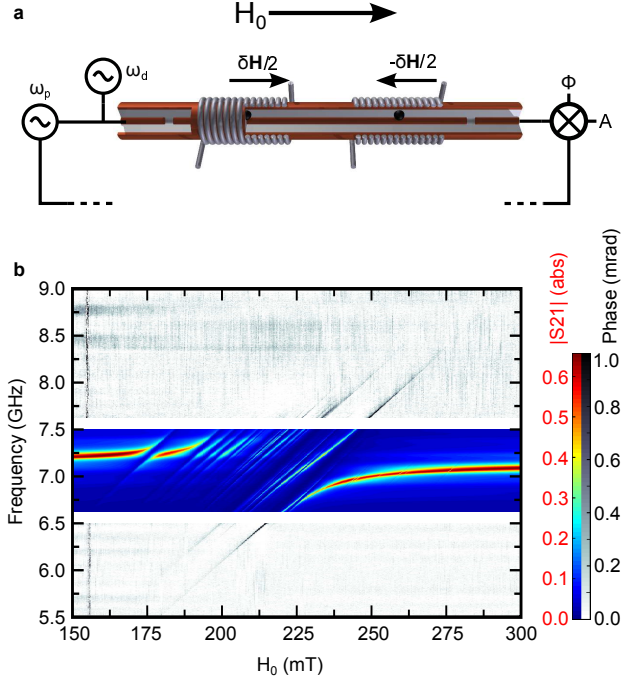}
\caption{Experimental scheme. (a) Experimental setup. Two YIG spheres are positioned at the magnetic field antinodes of the second harmonic of a transmission line cavity. Two sources, at $\omega_p$ and $\omega_d$, are coupled into the cavity. The transmitted amplitude and dispersive phase shift at $\omega_p$ is measured by homodyne detection. A global field, $H_0$, is applied to align the magnetisation of the spheres in the propagation direction of the cavity, and tune the ferromagnetic resonance to be off-resonance with the cavity modes. The field at each sphere is adjusted by an amount $\pm \delta H/2$ using a local coil wrapped around the cavity. (b) Response of the cavity as measured by transmission measurements of $\left|\textrm{S21}\right|$ (with $\omega_p$ between 6.625 GHz and 7.5 GHz and the source at $\omega_d$ turned off - coloured data) and dispersive measurements (with $\omega_p = \omega_1$, and $\omega_d$ between 5.5 GHz and 6.5 GHz, and 7.625 GHz and 9 GHz - greyscale data).}
\end{center}
\end{figure}

Having characterised the resonant coupling between the magnets and the cavity, we now move into the dispersive regime by adjusting $H_0$ such that the magnetostatic modes are significantly detuned from the cavity modes. In order to probe the magnetisation dynamics of the magnets we use a dispersive measurement technique\cite{Schuster2007,Gleyzes2007}, in which we measure the phase $\phi$ of the transmitted probe signal at a resonant frequency of the cavity, $\omega_p = \omega_{1,2}$, while applying a second drive tone $\omega_d$ of power 27 dBm to excite FMR.  The measured change in phase when FMR is driven is due to the reduction in the sum of the longitudinal components of magnetisation of the spheres\cite{Haigh2015}. Fig. 1b (monochrome data) shows the dispersive measurement of the uniform FMR modes of each individual magnetic elements above and below $\omega_2$, with them still detuned from each other by $\delta H = 16$ mT. Using this technique, we are able to measure FMR far detuned from the cavity modes.

In order to investigate the magnet-magnet coupling, $H_0$ is fixed such that the FMR frequencies of the uncoupled magnets ($\omega_{\textrm{F1}}$, $\omega_{\textrm{F2}}$) are $\approx 0.8$ GHz below $\omega_2$, and $\omega_p$ is set to $\omega_1$ to avoid resonant interactions between higher order FMR modes and the measurement cavity mode. We now attempt to bring the two magnetic modes through resonance by varying $\delta H$ from $-6$ mT to $+6$ mT. An avoided crossing is observed (Fig.~2a) demonstrating coupling of the magnetisation dynamics of the two spheres. The magnitude of the coupling is given by the frequency splitting of the modes at $\delta H=0$, giving a value of $2J/2 \pi=87$ MHz for this cavity detuning.

\begin{figure}
\begin{center}
\includegraphics{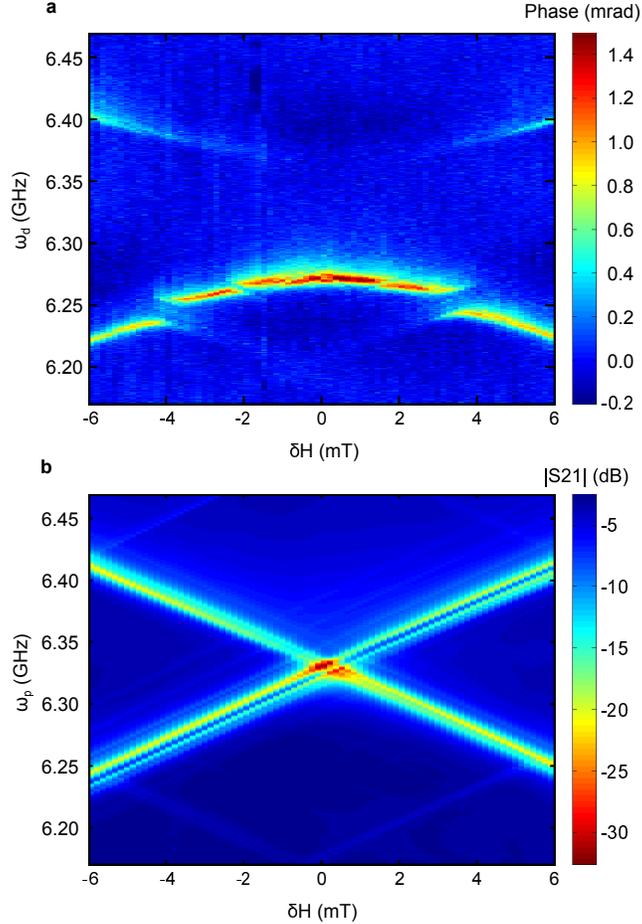}
\caption{Magnet-magnet coupling with and without a cavity, at $H_0=200$ mT. (a) Dispersively measured spectroscopy of the magnetostatic modes in the presence of a cavity close to and below $\omega_2$. An anticrossing between the modes is observed. The additional structure is due to weaker coupling to non-uniform magnetostatic modes. (b) Transmission amplitude $|$S21$|$ of the system with the coupling gaps closed, and therefore no cavity present. No anticrossing is seen, with the magnetostatic linewidth of $5$ MHz placing an upper bound on the magnet-magnet coupling.}
\end{center}
\end{figure}

We first exclude the possibility that this frequency splitting is simply due to the free space dipole coupling between the magnets. By closing the coupling gaps at both ends of the cavity, and reverting it back to a simple coaxial cable, the cavity modes are eliminated but the spatial separation of the magnets, and therefore the direct dipolar coupling, remains the same. In Fig.~2b we show $|$S21$|$ of the resulting transmission line as a function of $\delta H$, and observe the magnetostatic modes of the two spheres come into resonance with each other. This demonstrates that the splitting requires the cavity to be present, and is therefore not due to direct dipolar coupling. In addition, a simple calculation of the dipolar coupling can be made by linearizing the Landau-Lifshitz equation for the two individual magnets and including the dipolar stray field from each at the other. The harmonic solutions of the resulting coupled equations give a coupling rate of $2J = \gamma\frac{m_s}{2 \pi \left|r\right|^3}$. For YIG with a saturation magnetization of $m_s = 140$\,kA~m${^{-1}}$ and a magnet-magnet separation of $r=14$ mm, and setting the gyromagnetic ratio to $\gamma/2\pi=28$ GHz~T$^{-1}$, we obtain a value of $\approx 200$ kHz, much less than the coupling we observe.

The strong magnet-magnet coupling we observe in the presence of the cavity is analogous to that between qubits embedded in a microwave cavity\cite{Majer2007}, which is generally described using the Jaynes-Cummings Hamiltonian\cite{Jaynes1963}. Here, we adopt a quantum mechanical model of the Kittel mode of each magnet in the macrospin approximation \cite{Tavis1968,Soykal2010, Soykal2010a}. The magnetisation of the first (second) magnet is modelled by a large quantum angular momentum $S_{1(2)}$. The full Hamiltonian for a system with two magnetic spheres in the cavity is then

\begin{equation}
\begin{aligned}
\mathcal{H} = \gamma (H_0+\delta H/2) S_{1z}+ \gamma (H_0-\delta H/2) S_{2z}\\
+\gamma \delta H_v (a+a^\dagger)(S_{1x}\pm S_{2x}) + \hbar \omega_c a^\dagger a.
\end{aligned}
\end{equation}

This is the sum of the Zeeman energy of the two magnets in the total magnetic field including the cavity mode with r.m.s~vacuum magnetic field $\delta H$, and the photon energy of the cavity field with frequency $\omega_c$ and lowering operator $a$. The YIG spheres are located at antinodes of the cavity field and the a.c.~magnetic field lies along the x-direction, with the sign in the third term depending on the relative phase of a.c.~field at the two spheres.

We now move into an interaction picture with respect to the FMR resonance frequency of the Kittel mode. For simplicity we consider $\delta H=0$, and we define $\Delta = \omega_c-\gamma H_0$ to be the detuning of the cavity resonance from the FMR frequency. We also make the rotating wave approximation and obtain the interaction picture Hamiltonian

\begin{equation}
\mathcal{H}_{\rm int} = \hbar g_0 [a(S_{1+}\pm S_{2+})+a^\dagger(S_{1-}\pm S_{2-})] + \hbar \Delta a^\dagger a,
\end{equation}
where we have defined the single spin coupling frequency $g_0=\gamma \delta H_v$.

In the dispersive limit where $\Delta \gg g_0$ one can readily obtain, by applying second order degenerate perturbation theory to $\mathcal{H}_{\rm int}$, the effective Hamiltonian

\begin{equation}
\begin{aligned}
\mathcal{H}_{\rm eff} = &\pm \frac{2g_0^2}{\hbar\Delta}(S_{1x}S_{2x}+S_{1y}S_{2y})\\
&+ \frac{g_0^2}{\hbar\Delta}[S_{1x}^2+S_{1y}^2+S_{2x}^2+S_{2y}^2+\hbar(S_{1z}+S_{2z})]\\
&+ \frac{2g_0^2}{\Delta}a^\dagger a (S_{1z}+S_{2z}) +\Delta a^\dagger a.
\label{eq:model}
\end{aligned}
\end{equation}

This is perfectly analogous to the more familiar case of two qubits in a single cavity~\cite{Blais2004}. In the first term we find a coupling of the transverse components of the two magnetic moments, as observed in the experiment. The second term describes both linear and nonlinear shifts in the FMR resonance frequency and the third term is the dispersive coupling of the Kittel modes to the cavity which permits our dispersive measurement. The final term is the cavity detuning.

In the case where each YIG sphere is highly polarised it is usual to analyse this model in terms of the Holstein-Primakoff transformation~\cite{Holstein1940}, in which the magnetisation of the YIG sphere can be described in terms of a harmonic oscillator mode with lowering operator $b$ for which $S_z=\hbar(N_{\rm eff}/2-b^\dagger b)$, where $N_\textrm{eff}$ is the number of unpaired spins per magnet. Writing the first term of  equation (\ref{eq:model}) in terms of $b$ we find the coupling Hamiltonian $\mathcal{H}_c = \pm \frac{2g_0^2}{\hbar\Delta}(S_{1x}S_{2x}+S_{1y}S_{2y}) \approx \hbar J (b_2^\dagger b_1 + b_1^\dagger b_2)/2 $ where $J=\pm 2 N_{\rm eff} g_0^2/\Delta$. The expected normal mode splitting resulting from the coupling $J$ is dependent on the total coupling $g_{\omega_n}= g_0\sqrt{N_\textrm{eff}}$ of the magnetostatic mode to the $n$th cavity mode and inversely proportional to the detuning $\Delta$.

In order to verify that this is the correct coupling mechanism, we now study the dependence of $J$ on $\Delta$. We measure $J$ over a large range of FMR frequencies on both sides of the $\omega_2$ mode of the cavity (Fig. 3), whilst remaining in the dispersive regime. Although in Eqn \ref{eq:model} we have explicitly discussed only the effect of a single cavity mode it is straightforward to add other cavity modes to the model to obtain the formula for the observed normal mode splitting. In the measurement range there are three modes that play an important role in the coupling. When the coupling is dominated by a single mode, we observe the predicted $1/\Delta$ dependence, as previously observed in circuit QED\cite{Filipp2011}. We fit to this data the sum of the couplings due to the three modes, with corresponding magnet-cavity mode couplings, $g_{\omega_n}$, as free parameters. Our model is in good agreement with the data, and we extract values of $g_{\omega_1}/2 \pi=88$ MHz, $g_{\omega_2}/2 \pi=177$ MHz and $g_{\omega_3}/2 \pi=83$ MHz, consistent with the resonant couplings obtained from transmission measurements.

\begin{figure*}
\begin{center}
\includegraphics{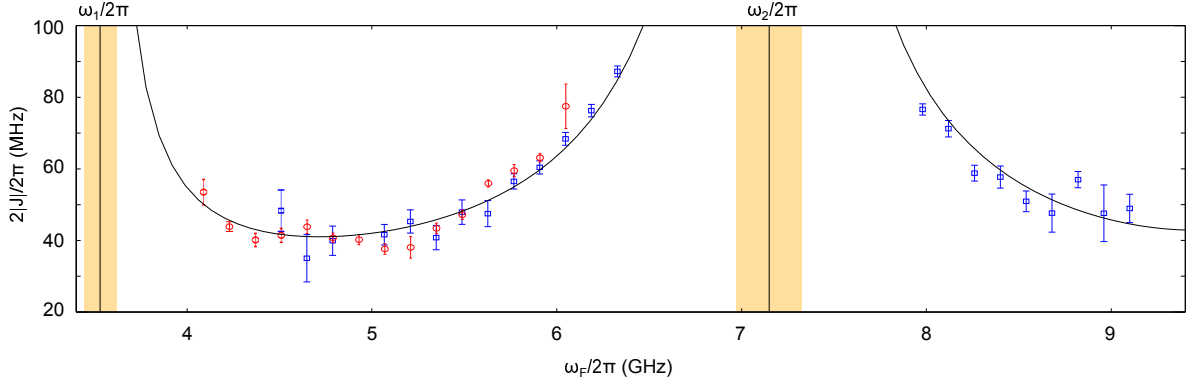}
\caption{Absolute magnet-magnet coupling, $2|J|/2 \pi$, as a function of FMR frequency of the uncoupled modes; this is determined by $H_0$. Blue squares are measured with $\omega_p=\omega_1$, and red circles are measured with $\omega_p=\omega_2$. Orange regions denote the resonant strong coupling regime\cite{Schuster2007}, where $\Delta<g_{\omega_n}$. The curve is from the model described in the text summed over the lowest three cavity modes, with values of magnet-cavity couplings as fitting parameters. We extract couplings of $g_{\omega_1}/2 \pi=88$ MHz, $g_{\omega_2}/2 \pi=177$ MHz and $g_{\omega_3}/2 \pi=83$ MHz, in approximate agreement with the coupling deduced from transmission measurements.}
\end{center}
\end{figure*}

The sign of $J$ is governed by the sign of the detuning and the spatial symmetry of the coupling modes, and determines whether the lowest energy state of the coupled system is symmetric ($S_{1x}=S_{2x}$) or antisymmetric ($S_{1x}=-S_{2x}$). These modes correspond to in phase and out of phase precession of the two magnetisations. In Figs 4a and 4b, we show two measured anticrossings above and below $\omega_2$. In both cases one of the eigenmodes cannot be driven; these dark states are similar to those seen in coupled superconducting qubits\cite{Filipp2011}. They are due to the microwave drive in the cavity forming a standing wave; at the sites of the two magnets the amplitude is in general different, and the phase difference can be either $0$ or $\pi$. If this does not match the the symmetries of the coupled FMR mode, then the mode cannot be driven and is dark.  In Figs 4c and 4d we show the predicted response of the system to a drive tone, based on the calculated form of the drive field in the cavity, the symmetry of the coupled mode as a function of $\delta H$ and the measured dispersive lineshape of the uncoupled modes. The visibility of the modes in the calculated response agrees well with our data.

\begin{figure}
\begin{center}
\includegraphics{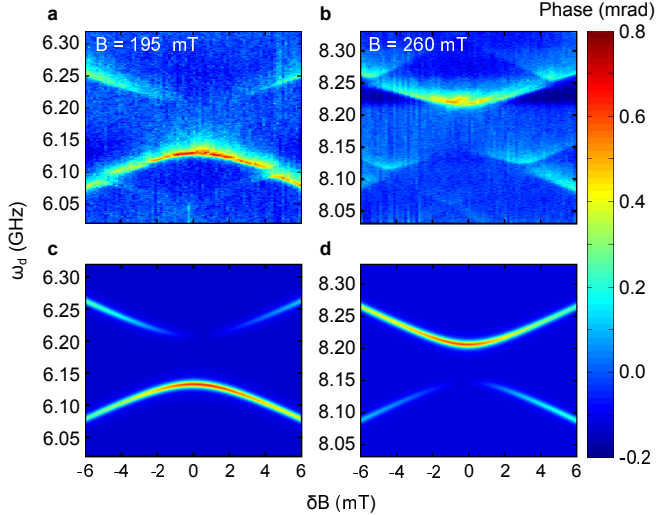}
\caption{Dark states and eigenmode symmetries. (a) Dispersively measured spectroscopy of the anticrossing between magnon modes below $\omega_2$ and (b) above $\omega_2$. (c) and (d) Modelled spectroscopy for \textbf{a} and \textbf{b}, taking into account detuning from the cavity modes and the frequency dependent drive symmetry, but not including non-uniform magnetostatic modes.}
\end{center}
\end{figure}

In conclusion, by measuring avoided crossings of magnetostatic modes we have shown that spatially separated magnetic moments may be passively coupled over a long range via dispersive coupling to the modes of an electromagnetic cavity. The coupling can be understood within the framework of circuit QED, putting magnets on a similar basis to qubits and atoms in cavities. This suggests a route towards coupling magnets to a variety of physical objects including mechanical oscillators and superconducting qubits using the dispersive strong coupling regime of QED, a different approach to that demonstrated by Tabuchi \textit{et al}\cite{Tabuchi2014a}. Such an approach might be used in coherent magnetic metamaterials, or to phase-lock many spatially separated magnetic oscillators, such as those in spin-torque nano-oscillators\cite{silva2008}. In such a scheme, the dispersive nature of the interaction means the linewidth of the coupled oscillators would not be limited by the quality of the cavity. Indeed, the linewidth of the cavity can be greater than the coupling rate between the magnets\cite{Majer2007}.

\begin{acknowledgments}
The authors would like to acknowledge discussions with Dr Andreas Nunnenkamp, which have contributed to the manuscript. We would like to acknowledge support from Hitachi Cambridge Laboratory, EPSRC Grant No. EP/K027018/1 and ERC grant 648613. AJF is supported by a Hitachi Research Fellowship. ACD is supported by the ARC via the Centre of Excellence in Engineered Quantum Systems (EQuS), project number CE110001013.
\end{acknowledgments}

\bibliographystyle{apsrev}
\bibliography{TwoSphereCoupling}

\end{document}